\documentclass[%
 reprint,
 amsmath,amssymb,
 aps,
]{revtex4-2}

\usepackage{graphicx}
\usepackage{dcolumn}
\usepackage{bm}

\begin{document}

\preprint{APS/123-QED}

\title{Best Annealing Path of Simulated Quantum Annealing\\
        via Efficient Adiabatic Phase Transition}
\author{Kiyotaka Murashima}
\email{murashima-kiyotaka@sei.co.jp}
\affiliation{%
 Nissin-Sumiden Energy System R$\textsl{\&}$D Center \\
 Sumitomo Electric Industries, Ltd., Kyoto, Japan
}%

\date{\today}

\begin{abstract}

Nested Simulated Annealing (NSA) is proposed for accelerating the classical simulation of Quantum Annealing (QA). 
NSA is based on two-spin interaction and exploits the causality of spin-flip, 
thereby significantly reducing computational cost.
And through Unitary Coupled-Cluster (UCC) method,
I establish a connection between NSA and $X\hspace{-1pt}X$ interaction
which is one of non-stoquastic Hamiltonians
and demonstrate that NSA effectively utilizes a trivial degenerate state.
Furthermore, by treating spins as continuous variables,
I derive an energy formulation that incorporates
both the problem Hamiltonian and the perturbation Hamiltonian
induced by transverse electromagnetic field
and show that the annealing path characterized by the lowest local-maximum exhibits the fastest convergence.
Based on the result, I present a strategy for finding optimal annealing parameters for Simulated Quantum Annealing (SQA).

\end{abstract}

\maketitle

\section{\label{sec:level1}Introduction}

Optimization problems arise in various domains, and Quadratic Unconstrained Binary Optimization (QUBO) is one of the formulations used to solve them~\cite{Glover}.
Since the Ising spin model can be represented in a similar form, Quantum Annealing (QA) has attracted considerable attention~\cite{VolksWagen}.
Furthermore, D-Wave has successfully commercialized QA systems.

QA was first proposed by T. Kadowaki and H. Nishimori~\cite{Kadowaki_1998}, and Path-Integral Monte Carlo is one of the fundamental approaches used to simulate its behavior~\cite{Martonak_2002}.
In this approach, the system Hamiltonian consists of two terms corresponding to longitudinal and transverse electromagnetic fields.
However, because these terms do not commute with each other, the simulation is performed using Suzuki-Trotter decomposition with a finite time intervals.
As a result, multiple spin configurations, referred to as trotter layers, emerge, and many studies have focused on computing them more efficiently and accurately~\cite{Muthumala_2020}.

However, Simulated Quantum Annealing (SQA) is generally slower than Simulated Annealing (SA)~\cite{King} because SQA must handle multiple trotter layers.
To address this issue, I previously proposed a method in which the trotter layers could be processed in parallel and demonstrated that it converged faster than conventional SQA~\cite{Murashima_1}~\cite{Murashima_2}.
In particular, the convergence was significantly accelerated when the spins directly interacting with it were preferentially selected for subsequent updates.
It was hypothesized that this speedup originated from trajectories passing through local energy-minimum state.

In order to simulate SQA fast, $X\hspace{-1pt}X$-catalysts have been investigated~\cite {Wurburton_1}~\cite{Wurburton_2}.
Moreover, a related mechanism has been examined in the context of adiabatic phase transition
between the driver Hamiltonian and the problem Hamiltonian~\cite{Matthias}.
These studies suggested that the probability of reaching the ground state could be enhanced 
by annealing path that traversed local energy-minimum.

In this paper, the relationship between Nested Simulated Annealing (NSA) and adiabatic phase transition is investigated.
NSA is based on two-spin interaction and exploits the causality.
Since the Pauli $X$-operator flips a single spin, 
the $X\hspace{-1pt}X$ interaction can be interpreted as simultaneously flipping two spins.
Therefore, there may be a relationship between the two approaches.

To establish this connection and demonstrate their effectiveness, spins are treated as continuous variables, 
and the spin energy associated with both the longitudinal and transverse fields are calculated analytically.
Section $\rm{I}\hspace{-1.2pt}\rm{I}$ reviews the SQA algorithm based on binary spins 
and extends it to the continuous variables.
Sections $\rm{I}\hspace{-1pt}\rm{I}\hspace{-1pt}\rm{I}$ and $\rm{I}\hspace{-1pt}\rm{V}$ review 
the parallelization scheme for trotter layers and the NSA algorithm, respectively.
Section $\rm{V}$ presents new results demonstrating a clear relationship between convergence speed and local-maximum of the energy,
and Section $\rm{V}\hspace{-1pt}\rm{I}$ reveals that the $X\hspace{-1pt}X$ interaction effectively utilizes degeneracy
through Unitary Coupled-Cluster (UCC) method 
and discusses the potential applicability to other non-stoquastic Hamiltonians,
Finally, Section $\rm{V}\hspace{-1pt}\rm{I}\hspace{-1.2pt}\rm{I}$ summarizes the results.

\section{\label{sec:level1}Simulated Quantum Annealing}
\subsection{\label{sec:level2}Algorithm with Binary Spins}

The Hamiltonian of the Ising spin model consists of two terms: a diagonal term ($H_d$) and an off-diagonal term ($H_o$), which do not commute with each other~\cite{Martonak_2002}.

\small
\begin{align}
    H  = - (H_d + H_o) = - \sum_{i<j} J_{ij} \hat{\sigma}_{z}^{i} \hat{\sigma}_{z}^{j} - \sum_{i} \mathit{\Gamma} \hat{\sigma}_{x}^{i} ,
\end{align}
\normalsize

where $J_{ij}$ is the interaction coefficient between the $i$-th and $j$-th spins,
$\hat{\sigma}_{z}^{i}$ and $\hat{\sigma}_{z}^{j}$ are the Pauli $Z$-operators acting on the $i$-th and $j$-th spins, respectively,
and $\hat{\sigma}_{x}^{i}$ is the Pauli $X$-operator acting on the $i$-th spin.
The magnitude of the transverse electromagnetic field is denoted by $\mathit{\Gamma}$.
Therefore, the optimization problem is to find the spin configuration that maximizes the partition function $Z$, below:

\begin{figure}[t!]
\includegraphics[keepaspectratio,scale=0.36, bb=-65 0 700 660]{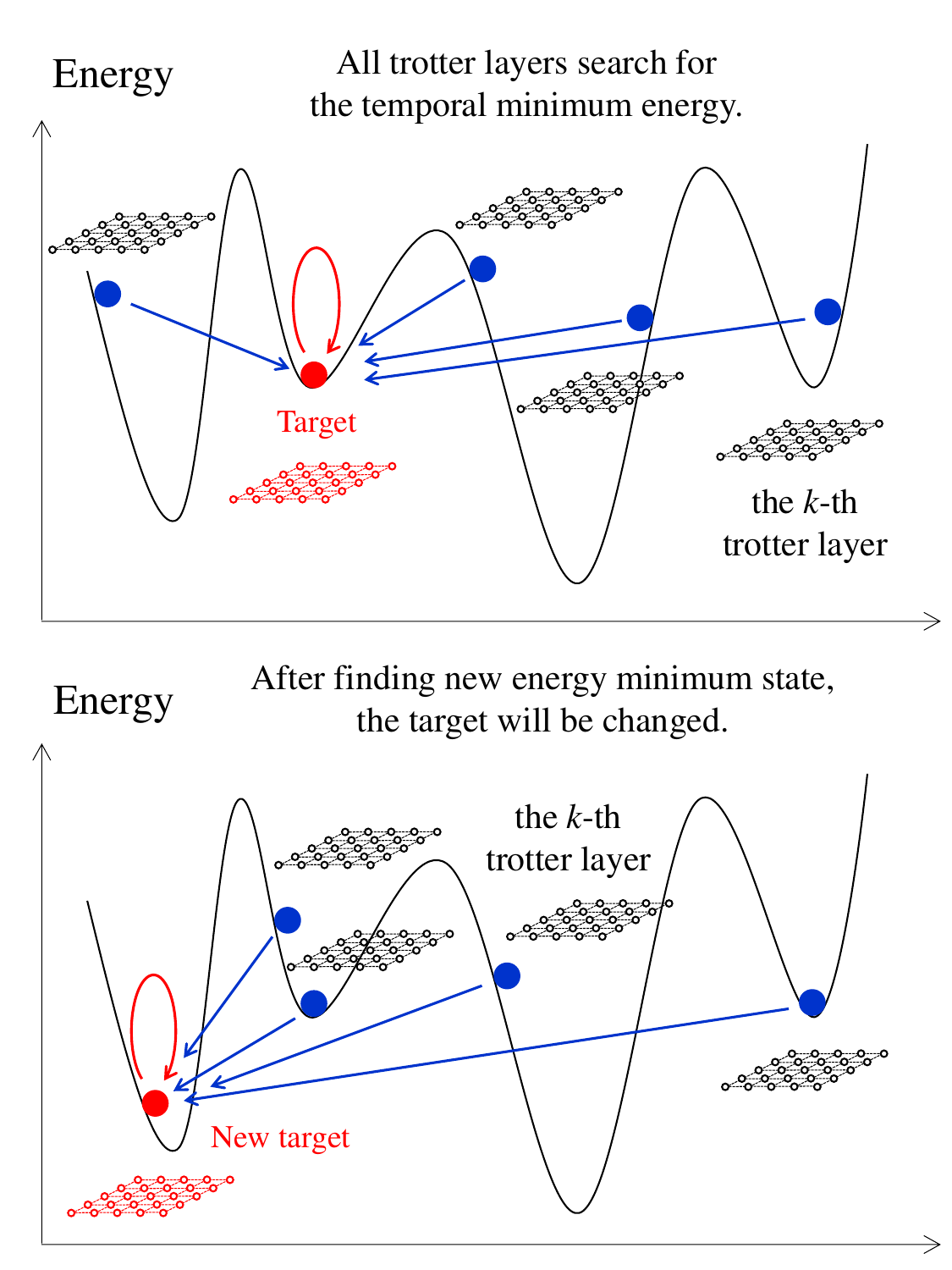}
\caption{
A schematic illustration of search process for the minimum energy is shown.
All $P$ trotter layers update their spin configurations independently,
whereas conventional SQA must satisfy a periodic boundary condition.
The transition probability is determined according to the Metropolis criterion,
and the second term in Eq.~(6) acts as a penalty term
when the $i$-th spin is not the same spin direction as that of the temporal energy-minimum state.
Once a new minimum energy is found, the target state is updated immediately.
}
\label{fig:concept}
\end{figure}

\small
\begin{align}
    Z  = \mathrm{Tr} ( e^{-\beta H} )
        = \sum_{\psi_0} \langle \psi_0 | e^{-\beta H} |\psi_0 \rangle.
\end{align}
\normalsize

$|\psi_0\rangle$ is the wavefunction of the system,
and $\beta = 1/k_{B}T$, where $k_B$ is the Boltzmann constant and $T$ is the temperature.

After applying Suzuki--Trotter decomposition and inserting Identity Operators, we obtain

\small
\begin{align}
    Z  &= \langle \psi_0 | e^{-\beta (\frac{H_d}{P} + \frac{H_o}{P})^{P}} |\psi_0 \rangle \nonumber\\
        &= \hspace{-0.3cm} \sum_{\psi_0, \cdot \cdot \cdot ,\psi_{2P-1}}
              \langle \psi_0 | e^{-\beta \frac{H_d}{P}} |\psi_{2P-1} \rangle
              \langle \psi_{2P-1} | e^{-\beta \frac{H_o}{P}} |\psi_{2P-2} \rangle \nonumber\\
             &  \ \ \ \ \ \ \ \ \cdot \cdot \cdot \cdot \cdot \cdot \ \ 
              \langle \psi_2 | e^{-\beta \frac{H_d}{P}} |\psi_1 \rangle
              \langle \psi_1 | e^{-\beta \frac{H_o}{P}} |\psi_0 \rangle \nonumber\\
        &= \hspace{-0.3cm} \sum_{\psi_0, \cdot \cdot \cdot ,\psi_{P-1}}
              \langle \psi_0 | e^{-\beta \frac{H_d}{P}} |\psi_0 \rangle
              \langle \psi_0 | e^{-\beta \frac{H_o}{P}} |\psi_{p-1} \rangle \nonumber \\
             &  \ \ \ \ \ \ \ \ \cdot \cdot \cdot \cdot \cdot \cdot \ \  
              \langle \psi_1 | e^{-\beta \frac{H_d}{P}} |\psi_1 \rangle
              \langle \psi_1 | e^{-\beta \frac{H_o}{P}} |\psi_0 \rangle .
\end{align}
\normalsize

In the last expression, the fact is used that $e^{-\beta \frac{H_d}{P}}$ is a diagonal operator.
By carrying out further calculations, the energy of the system can be finally represented in a classical form~\cite{Martonak_2002}.

\small
\begin{align}
    E =& \sum_{k=0}^{P-1} \left (
                             \sum_{i<j} J_{ij} s_{z}^{i, k} s_{z}^{j, k} + \sum_{i} J_{\perp} s_{z}^{i, k} s_{z}^{i, k+1}
                             \right ) , \\ 
    &  \hspace{0.5cm} \ \ \ \Biggl ( J_{\perp} = - \frac{P T}{2} \mathrm{ln} \left (
                                                        \mathrm{tanh} \frac{\mathit{\Gamma}} {P T}
                                                        \right )  \Biggr ) \nonumber
\end{align}
\normalsize

\begin{figure}[t!]
\includegraphics[keepaspectratio,scale=0.6, bb=-45 0 365 350]{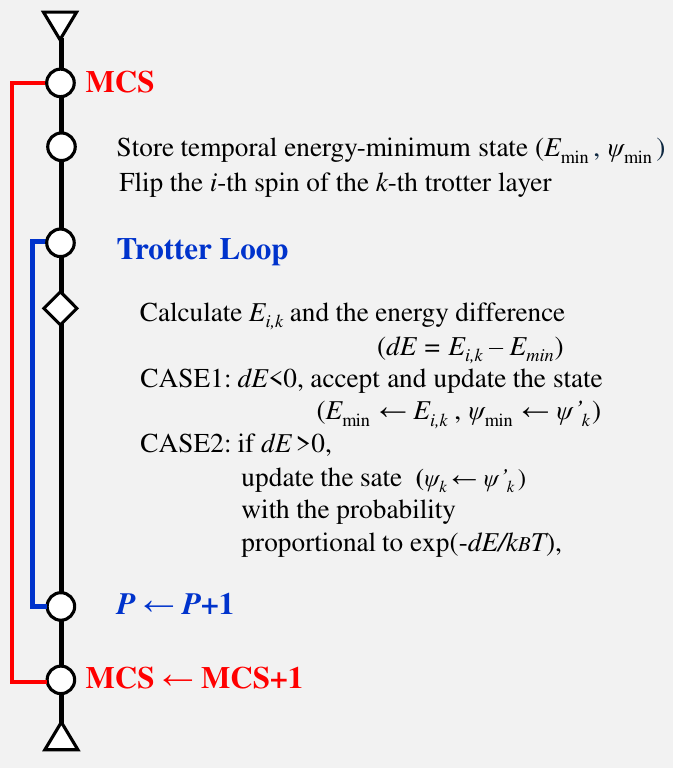}
\caption{
A flowchart of the trotter layer parallelization algorithm is shown.
In conventional SQA, the energies of all trotter layers must be summed, which hinders rapid convergence.
The proposed algorithm avoids this procedure, and all trotter layers mimic the spin configuration
corresponding to the temporal energy-minimum state.
By sharing such energy information, each trotter layer keeps searching for its global minimum energy independently.
Such an quantum effect is preserved, resulting in faster convergence.
}
\label{fig:flowchart}
\end{figure}

where $s_{z}^{i,k}$ denotes the classical $i$-th spin,
taking values in $\{\pm 1\}$, on the $k$-th trotter layer.
As shown in Eq.~(4), the Hamiltonian of the two-dimensional Ising model turns out a three-dimensional
with an additional axis corresponding to the imaginary time.
The trotter layers form torus and must satisfy the periodic boundary condition $ s_{z}^{i,0} = s_{z}^{i,P} $.
Under this constraint, it is necessary to find the spin configuration that minimizes the spin energy.

From this point onward, I will rewrite Eq.~(4) in a quantum manners.

\small
\begin{align}
    E = \sum_{k=0}^{P-1}  \ \biggl (&
                              \langle \psi_k |
                              \sum_{i<j} J_{ij} \hat{\sigma}_{z}^{i, k} \hat{\sigma}_{z}^{j, k}
                              | \psi_k \rangle \nonumber \\
       & \ \ \ +
                              \langle \psi_k |
                              \sum_{i} J_{\perp} \hat{\sigma}_{z}^{i, k} \hat{\sigma}_{z}^{i, k+1}
                              | \psi_{k+1} \rangle
                              \biggl ).
\end{align}
\normalsize

Since the objective is to find the energy-minimum state, it is reasonable to perform the search in step by step.
Therefore, the order of summing the energies over all trotter layers and finding the spin configuration
that minimizes the total energy can be exchanged.
First, the temporal energy-minimum state is determined at every Monte Carlo Step (MCS).
This implies that $|\psi_{k+1}\rangle$ can be replaced by ${\mathrm{argmin}} |\psi_{k+1}\rangle$. 
Finally, the global energy-minimum state can be obtained according to Eq.~(6).

\small
\begin{align}
    & \hspace{-1.5mm}  \mathrm{argmin}  \ E \nonumber \\
    & \hspace{-1.5mm} = \underset{\psi_{k}}{\mathrm{argmin}} \ 
                              P \times \biggl (
                              \langle \psi_{k} |
                              \sum_{i<j} J_{ij} \hat{\sigma}_{z}^{i, k} \hat{\sigma}_{z}^{j, k}
                              | \psi_{k} \rangle \nonumber \\
       &  \ \ \ \ \ \ \ \ \ \ \ +
                              \langle \psi_{k} |
                              \sum_{i} J_{\perp} \hat{\sigma}_{z}^{i, k} \hat{\sigma}_{z}^{i, \mathit{min}}
                              \Bigl{\{} \underset{\psi_{k+1}}{\mathrm{argmin}} \ | \psi_{k+1} \rangle \Bigl{\}}
                              \biggl )   \nonumber \\
    & \hspace{-1.5mm} = \underset{\psi_{k}}{\mathrm{argmin}} \ 
                              P \times \biggl (
                              \langle \psi_{k} |
                              \sum_{i<j} J_{ij} \hat{\sigma}_{z}^{i, k} \hat{\sigma}_{z}^{j, k}
                              | \psi_{k} \rangle \nonumber \\
       &  \ \ \ \ \ \ \ \ \ \ \ \ \ \ \ \ \ \ \ \ \ \ \ \ \ +
                              \langle \psi_{k} |
                              \sum_{i} J_{\perp} \hat{\sigma}_{z}^{i, k} \hat{\sigma}_{z}^{i, \mathit{min}}
                              | \psi_{\mathit{min}} \rangle
                              \biggl ).
\end{align}
\normalsize

Equation~(6) indicates that the problem reduces to finding the global energy-minimum state
while allowing the $P$ candidate trotter layers to mimic the temporal energy-minimum state.
To describe this more clearly, $\hat{\sigma}_{z}^{i,\mathrm{min}}$ is introduced to denote the Pauli $Z$-operator
corresponding to the temporal energy-minimum state.
The concept of the algorithm and its flowchart are illustrated in Fig.~\ref{fig:concept}.
The quantum effect still remains through the second term, 
in which each trotter layer searches for the energy-minimum state independently.
Although this approach is not mathematically rigorous, we have demonstrated its effectiveness in my previous studies~\cite{Murashima_1,Murashima_2}.
It should be noted, however, that the algorithm omits a step derived from rigorous quantum mechanics.
As a result, certain exotic phenomena may be lost, such as the ``worldline''~\cite{Ammon_1997} 
and the ``sign problem''~\cite{Henelius_2000}.

\subsection{\label{sec:level2}Augmentation to Continuous Variables}

As long as the spin is treated as a binary variable,
the spin energy along the $x$ direction is always zero, even when the transverse electromagnetic field is nonzero.
Therefore, in order to evaluate the spin energy exactly, the spins must be treated as continuous variables.
In the following, I derive the total spin energy using continuous spin variables.
Before proceeding, let us slightly modify the Hamiltonian of the system.
A parameter $\lambda$ is introduced into the problem Hamiltonian, which is eventually set to one.

\small
\begin{align}
    H  &= - (\lambda H_{d} + H_o) 
          =- \lambda \sum_{i<j} J_{ij} \hat{\sigma}_{z}^{i} \hat{\sigma}_{z}^{j} - \sum_{i} \mathit{\Gamma} \hat{\sigma}_{x}^{i}
\end{align}
\normalsize

In Eq.~(3), the Identity Operator, $\mathit{I} = |0\rangle\langle 0| + |1\rangle\langle 1|$, was inserted.
Here, the spin is treated as a continuous variable constrained to the $x$--$z$ plane.
it is sufficient to consider spin rotations about the $y$ axis on the Bloch sphere.
To this end, the rotation operator $\mathit{R}_y(\theta)=e^{-i\frac{\theta}{2}\hat{\sigma}_y}$ is introduced,
where $\hat{\sigma}_y$ is the Pauli $Y$-operator and $\theta$ is the azimuthal angle.
Therefore, the identity operator below is used.

\small
\begin{align}
    \mathit{I} &=  \mathit{R}_y (\theta) \bigl ( | 0 \rangle \langle 0| +  |1 \rangle \langle 1| \bigl ) \mathit{R}_y (-\theta) \nonumber \\
                  &=   
                    \left( \hspace{-0.15cm}
                        \begin{array}{ccc}
                            \cos{\frac{\theta}{2}} \\
                            \sin{\frac{\theta}{2}}
                        \end{array}
                    \hspace{-0.15cm} \right)
                    \left( \hspace{-0.15cm}
                        \begin{array}{ccc}
                            \cos{\frac{\theta}{2}}  \ \sin{\frac{\theta}{2}}
                        \end{array}
                    \hspace{-0.15cm} \right)
               +   
                    \left( \hspace{-0.15cm}
                        \begin{array}{ccc}
                            -\sin{\frac{\theta}{2}} \\
                            \cos{\frac{\theta}{2}}
                        \end{array}
                    \hspace{-0.15cm} \right)
                    \left( \hspace{-0.15cm}
                        \begin{array}{ccc}
                            -\sin{\frac{\theta}{2}}  \ \cos{\frac{\theta}{2}}
                        \end{array}
                    \hspace{-0.15cm} \right) \nonumber
\end{align}
\normalsize

Then, the partition function is proportional to the product of the following expressions:

\small
\begin{align}
%
                       & Z \varpropto  \langle \psi_{k+1} | e^{-\beta \frac{H_{d}}{P}}
                       \bigl ( \mathit{R}_y (\theta_{k})                      \bigl (  |+ \rangle\hspace{-0.05cm}+\hspace{-0.05cm}|- \rangle \bigl ) \nonumber \\ 
                       &\hspace{+2.0cm}                                       +  \bigl ( \langle -|\hspace{-0.05cm}+\hspace{-0.05cm}\langle +|  \bigl ) 
                                \mathit{R}_y (-\theta_{k}) \bigl )
                        e^{-\beta \frac{H_{o}}{P}} | \psi_{k-1} \rangle
\end{align}
\normalsize


The states $| + \rangle \hspace{-0.1cm} = \hspace{-0.1cm} \frac{|0\rangle+|1\rangle}{\sqrt{2}}$ and $| - \rangle \hspace{-0.1cm} = \hspace{-0.1cm} \frac{|0\rangle-|1\rangle}{\sqrt{2}}$ are adopted as a new basis.
Accordingly, the Identity Operator $ | 0 \rangle \langle 0| + |1 \rangle \langle 1| $ can be written as $\bigl(| + \rangle  + | - \rangle \bigr) \bigl( \langle -| + \langle +| \bigr) $.

If the system evolves adiabatically and the difference between $\theta_k$ and $\theta_{k\pm1}$ is very small,
the wavefunction of the $k$-th trotter layer can be expressed as
$| \psi_{k} \rangle \hspace{-0.1cm} = \hspace{-0.1cm} \frac{1}{\sqrt{2}} {\mathit{R}}_{y}(\theta_{k})
\bigl(| + \rangle + | - \rangle\bigr) \hspace{-0.02cm}$.
Under this condition, $|\psi_{k\pm1}\rangle$ is approximated as follows:

\small
\begin{align}
\begin{split}
         & | \psi_{k+1} \rangle =  \frac{1}{\sqrt{2}}  \bigl \{ {\mathit{R}}_{y}(\theta_{k})(| + \rangle\hspace{-0.05cm}+\hspace{-0.05cm}| - \rangle)
        +  {\mathit{R}}_{y}(\theta_{k+1}^{'})(| + \rangle\hspace{-0.05cm}+\hspace{-0.05cm}| - \rangle) \bigr \} \\
         & | \psi_{k-1} \rangle =  \frac{1}{\sqrt{2}} \bigl \{ {\mathit{R}}_{y}(\theta_{k})(| + \rangle\hspace{-0.05cm}+\hspace{-0.05cm}| - \rangle)
        -  {\mathit{R}}_{y}(\theta_{k-1}^{'})(| + \rangle\hspace{-0.05cm}+\hspace{-0.05cm}| - \rangle) \bigr \}
\end{split}
\end{align}
\normalsize

where $\theta_{k\pm1}'$ denotes the angular difference between $\theta_k$ and $\theta_{k\pm1}$.
Substituting Eq.~(9) into Eq.~(8) and ignoring a constant value, we obtain

\small
\begin{align}
                    & \hspace{0.2cm} Z \varpropto \bigl ( \bigl (\langle -|\hspace{-0.05cm}+\hspace{-0.05cm}\langle +| \bigl ) {\mathit{R}}_{y}^\mathsf{T}(\theta_{k})
                               + \bigl (\langle -|\hspace{-0.05cm}+\hspace{-0.05cm}\langle +| \bigl ) {\mathit{R}}_{y}^\mathsf{T}(\theta_{k+1}^{'})
                       \bigl ) \nonumber \\
                    & \hspace{1.0cm} e^{-\beta \frac{H_{d}}{P}}
                                \bigl ( \mathit{R}_y (\theta_{k})  \bigl (  |+ \rangle\hspace{-0.05cm}+\hspace{-0.05cm}|- \rangle \bigl )
                       \bigl ( \langle -|\hspace{-0.05cm}+\hspace{-0.05cm}\langle +|  \bigl ) 
                                \mathit{R}_y (-\theta_{k}) \bigl ) \nonumber \\
                    & \hspace{1.0cm} e^{-\beta \frac{H_{o}}{P}}
                                \bigl ( \mathit{R}_y (\theta_{k})  \bigl (  |+ \rangle\hspace{-0.05cm}+\hspace{-0.05cm}|- \rangle \bigl )
                              - \mathit{R}_y (\theta_{k-1}^{'})  \bigl (  |+ \rangle\hspace{-0.05cm}+\hspace{-0.05cm}|- \rangle \bigl ) \bigl ).  \nonumber
\end{align}
\normalsize

Since $H_d$ is a diagonal matrix and $H_o$ is an off-diagonal,
the partition function can be calculated as

\small
\begin{align}
                   \hspace{0.3cm} Z &\varpropto 
                   \bigl (\langle -| \hspace{-0.05cm}+\hspace{-0.05cm}\langle +| \bigl ) {\mathit{R}}_{y}^\mathsf{T}(\theta_{k})
                              \hspace{+0.05cm} e^{-\beta \frac{H_{d}}{P}}
                    \mathit{R}_y (\theta_{k})  \bigl (  |+ \rangle \hspace{-0.05cm}+\hspace{-0.05cm}|- \rangle \bigl ) \nonumber \\
                    &+ 
                   \bigl ( \langle -| \hspace{-0.05cm}+\hspace{-0.05cm}\langle +|  \bigl ) \mathit{R}_y (-\theta_{k})
                              \hspace{+0.05cm} e^{-\beta \frac{H_{o}}{P}}
                    {\mathit{R}}_{y}(\theta_{k}) \bigl (|+ \rangle \hspace{-0.05cm}+\hspace{-0.05cm}|- \rangle \bigl )  \nonumber \\
                    &- 
                    \bigl ( \langle -| \hspace{-0.05cm}+\hspace{-0.05cm}\langle +|  \bigl )
                              \mathit{R}_y (-\theta_k)
                              \hspace{+0.05cm} e^{-\beta \frac{H_{o}}{P}}
                    {\mathit{R}}_{y}(\theta_{k-1}^{'}) \bigl (|+ \rangle \hspace{-0.05cm}+\hspace{-0.05cm}|- \rangle \bigl ).
\end{align}
\normalsize

Using ${R}_y^{\mathsf{T}}(\theta_k) = {R}_y(-\theta_k)$, the first and the second terms in Eq.~(10)
can be expressed in a form analogous to the classical spin energy~\cite{Smolin,Shin}.
However, the quantum effect remains in the last term of Eq.~(10) as a penalty term.
Each term can be calculated as below:

\small
\begin{align}
    \hspace{0.3cm}
    & \langle \psi (\theta_k) | e^{-\beta \frac{H_{d}}{P}} |\psi (\theta_k) \rangle
    + \langle \psi (\theta_k) | e^{-\beta \frac{H_{o}}{P}} |\psi (\theta_k) \rangle \nonumber \\
        &
        = \exp
           { \left (                   - \frac{\lambda \sum_{i<j}               J_{ij}                   \cos \theta_k^{i}                  \cos \theta_k^{j}
                + \sum_{i}  \mathit{\Gamma}   \sin \theta_k^{i} }{P T}  
                     \right ) } \\
\nonumber \\
    & \langle \psi(\theta_k) | e^{-\beta \frac{H_{o}}{P}} |\psi(\theta_{k-1}^{'}) \rangle \nonumber \\
        & 
          = \biggl \langle \psi(\theta_k) \biggl | \exp \left( {-\frac{\sum_{i} \mathit{\Gamma} \hat{\sigma}_{x}^{i}}{P T}} \right) \biggl | \psi(\theta_{k-1}^{'}) \biggl \rangle \nonumber \\
        & 
          = \biggl \langle \psi(\theta_k) \biggl | \prod_{i} \biggl ( \cosh \biggl (\frac{\mathit{\Gamma}}{P T} \biggl) + \hat{\sigma}_{x}^{i} \sinh \biggl (\frac{\mathit{\Gamma}}{P T} \biggl ) \biggl )  \biggl | \psi(\theta_{k-1}^{'}) \biggl \rangle \nonumber \\
        & 
        = \prod_{i} 
        \left( \hspace{-0.1cm}
            \begin{array}{ccc}
                \cos{\frac{\theta_{k}^i}{2}} \\
                \sin{\frac{\theta_{k}^i}{2}}
            \end{array}
        \hspace{-0.1cm} \right)^{ \hspace{-0.1cm} \mathsf{T} }\hspace{-0.2cm}
        \left( \hspace{-0.1cm}
            \begin{array}{ccc}
                \cosh(X) & \sinh(X) \\
                \sinh(X) & \cosh(X)
            \end{array}
        \hspace{-0.1cm} \right)  \hspace{-0.1cm}
        \left( \hspace{-0.1cm}
            \begin{array}{ccc}
                \cos{\frac{\theta_{k-1}^{i'}}{2}} \\
                \sin{\frac{\theta_{k-1}^{i'}}{2}}
            \end{array}
        \hspace{-0.1cm} \right) \nonumber \\
        & 
         = \prod_{i} 
             \frac{1}{2} \{\sinh(2X) \}^\frac{1}{2}
             \biggl \{
             e^{X} \cos \biggl( \frac{\theta_k^i}{2} - \frac{\theta_{k-1}^{i'}}{2} \biggl) \nonumber \\
             & \hspace{3.5cm}
             + e^{-X} \sin \biggl( \frac{\theta_k^i}{2} + \frac{\theta_{k-1}^{i'}}{2} \biggl)
             \biggl \}
\end{align}
\normalsize

In Eq.~(12), $X$ is defined as $\frac{J_{\perp}}{PT}$, where $J_{\perp}$ is given in Eq.~(4).
Finally, the partition function of the system can be formulated as follows:

\small
\begin{align}
     &\hspace{-0.1cm}Z = \langle \psi_0 | e^{-\beta \frac{H}{T}} |\psi_0 \rangle \nonumber \\
     &\hspace{0.17cm} \varpropto \prod_{k=0}^{P-1} \exp
           { \left ( \hspace{-0.03cm} - \frac{\lambda \sum_{i<j} \hspace{-0.03cm}  J_{ij}  \hspace{-0.03cm} \cos \theta_k^{i} \hspace{-0.02cm}  \cos \theta_k^{j}
           \hspace{-0.03cm} + \hspace{-0.03cm}  \sum_{i} \hspace{-0.03cm} \mathit{\Gamma}  \sin \theta_k^{i} }{PT}  
                     \right ) } \nonumber \\
      &\ \ \ \ \ \times \prod_{i} 
             \frac{1}{2} \{\sinh(2X) \}^\frac{1}{2}
             \biggl \{
             e^{X} \cos \biggl( \frac{\theta_k^i}{2} - \frac{\theta_{k-1}^{i'}}{2} \biggl) \nonumber \\
      &\ \ \ \ \ \hspace{3.5cm} \ 
             + e^{-X} \sin \biggl( \frac{\theta_k^i}{2} + \frac{\theta_{k-1}^{i'}}{2} \biggl)
             \biggl \}  \nonumber
\end{align}
\normalsize

Following the reasoning of Eq.~(6), the problem reduces to finding the set of $\theta_k^{i}$ values
that minimizes the spin energy while allowing each $\theta_k^{i}$ to mimic $\theta_{\mathrm{min}}^{i}$,
where $\theta_{\mathrm{min}}^{i}$ denotes the angle of the $i$-th spin in the temporal energy-minimum state.
The final result can be reduced to the following equation:

\small
\begin{align}
          \hspace{-0.5cm}
          E &= \sum_{k=0}^{P-1} \biggl [
                             \lambda \sum_{i<j} J_{ij} \cos \theta_{k}^{i} \cos \theta_{k}^{j}
                             + \sum_{i} \mathit{\Gamma} \sin \theta_{k}^{i}   \nonumber \\
             &\hspace{+1.0cm} - PT \ln \biggl \{
                e^{X} \cos \biggl( \frac{\theta_k^i}{2} - \frac{\theta_{\mathit{min}}^i}{2} \biggl) 
                    \nonumber \\
             &\hspace{+2.4cm} + e^{-X} \sin \biggl( \frac{\theta_k^i}{2} + \frac{\theta_{\mathit{min}}^i}{2} \biggl)
             \biggl \}  
                                  \biggl ]
\end{align}
\normalsize

\begin{figure}[t!]
\includegraphics[keepaspectratio,scale=0.45, bb=55 0 500 327]{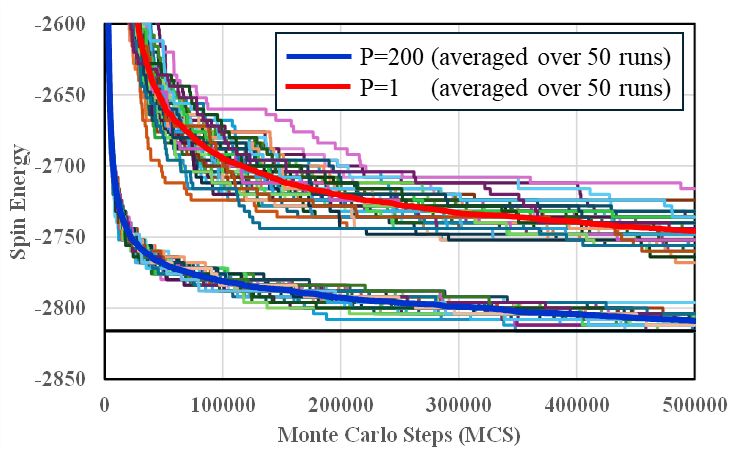}
\caption{
The results demonstrate that the trotter layer parallelization method
inspired by Quantum Monte Carlo (QMC) is effective.
Each trotter layer converges toward the global minimum energy,
while mimicking the temporal energy-minimum state.
The convergence curves are obtained by averaging over 50 runs.
Hereinafter, only the averaged results are presented.
}
\label{fig:G34_SQA}
\end{figure}

\section{\label{sec:level1}Trotter Layer Parallelization}

The validity of the proposed method is examined by solving GSET~\cite{GSET}.
GSET has been widely used to evaluate the performance of various optimization algorithms~\cite{Muthumala_2020,Benlic}, 
as its minimum energy have been extensively studied. 
The computations are performed using Python 3.10.7 on Intel$^{\circledR}$ Core$^{\mathrm{TM}}$ i5-10210U processor,
following the procedure shown in Fig.~\ref{fig:flowchart}.
The simulation parameters are summarized in TABLE~\rm{I}.
The temperature is fixed at sufficiently low to emulate realistic quantum annealing conditions.
The initial magnitude of $\mathit{\Gamma}$ is set to 1.0 and subsequently quenched.
The annealing schedule is shown by the red dotted line in Fig.~\ref{fig:G34_3} and Fig.~\ref{fig:G9_3}.
The spins are updated sequentially and flipped on each trotter layer.
The convergence curves are obtained by averaging the results over 50 runs for each simulation.

\begin{table}[h!]
  \small
  \caption{Graph Characteristics and Simulation Parameters}
  \label{table:data_type}
  \centering
  \begin{tabular}{c||c|c|c||c|c}
    \hline
    Graph  &\  Nodes  &\ Edges &\ $E_{\mathit{min}}$ &\ $P$ &\ Temp.  \\
    \hline
            G9 \  & \ \ 800   \ \    & \ \ 19,176  \ \ & \ \ -4,172 \ \  & \ \ 150 \ \ & \ \ 0.004  \\
            G34 \ & \ \ 2,000 \ \    & \ \ 4,000 \ \   & \ \ -2,816 \ \   & \ \ 200 \ \ & \ \ 0.001   \\
    \hline
  \end{tabular}
  \label{tb:param}
\end{table}

The results demonstrating the validity of parallelizing the trotter layers are shown in Fig.~\ref{fig:G34_SQA}.
The convergence behaviors of the proposed algorithm described by Eq.~(6)
are compared with SA for G34.
For conventional SA, the number of trotter layers, $P$, is set to one,
while the temperature schedule is kept identical to that used in the proposed algorithm.
Therefore, the transition probabilities are expected to be comparable to those of SQA.

As shown in Fig.~\ref{fig:G34_SQA}, the proposed algorithm reaches very close to the global minimum energy,
whereas the conventional SA appears to become trapped in a higher energy state before reaching it. 
This result suggests that the energy-minimum state of the Ising model can be found more efficiently
through the quantum-like effect of mimicking the temporal energy-minimum state.
Admittedly, the comparison is not entirely fair
because the proposed method performs approximately $P$ times more trials at each MCS.
Consequently, the computational cost per MCS is higher.
However, this drawback can be mitigated because the trotter layers can be processed in parallel.
By exploiting trotter layer parallelization, the proposed method can achieve a computation time
comparable to that of the conventional SA.

\section{\label{sec:level1}Nested Simulated Annealing}
\subsection{\label{sec:level2}Concept of NSA}

\begin{figure}[t!]
\includegraphics[keepaspectratio,scale=0.6, bb=-7 0 430 490]{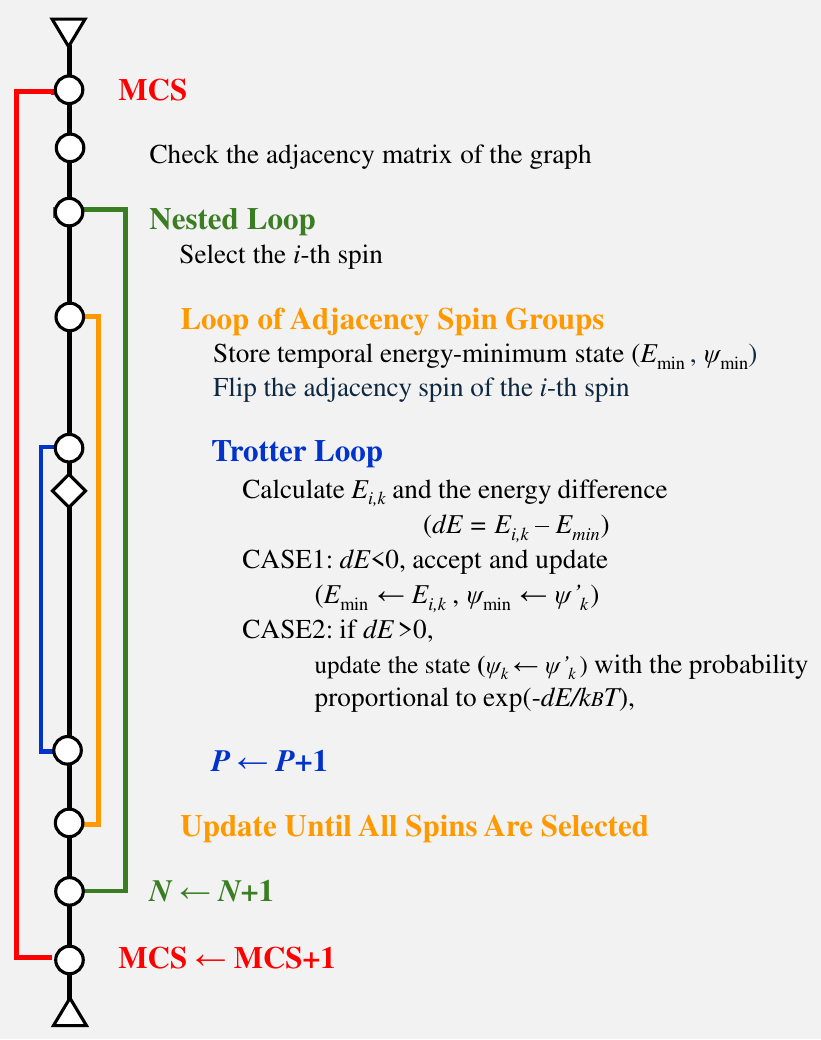}
\caption{
A flowchart of the NSA algorithm is shown.
Spins are selected from groups determined by the adjacency matrix of the graph, 
and the spins are flipped one by one on each trotter layer.
Compared with the algorithm shown in Fig.~\ref{fig:flowchart}, the green and yellow loops have been added.
By repeatedly executing these nested loops in each Monte Carlo Step (MCS),
the neighboring spin groups converge to their local minimum.
}
\label{fig:flowchart2}
\end{figure}

When the trotter layers are treated independently,
a large number of replicas must be prepared, as in ``parallel tempering''~\cite{Wang},
although computational speedup can be achieved.
This approach requires substantial computational resources, particularly memory capacity.
To address this issue, I proposed the NSA.
NSA is motivated by the observation that spin relaxation preferentially occurs among neighboring spins
and that neighboring spin groups tend to converge to their local-minimum.
Furthermore, because there are $P$ candidate trotter layers,
the acceleration of convergence can be accumulated across the search process.
A flowchart of the algorithm is shown in Fig.~\ref{fig:flowchart2}.
The effectiveness of NSA is evaluated using both the sparse graph G34 and the complex graph G9.
For ease of comparison, the total number of inner loop per MCS, $N \hspace{-1.2pt} \times \hspace{-1.2pt} P$,
is fixed at 200 for G34 and 150 for G9.
However, an additional loop, shown as the yellow loop in Fig.~\ref{fig:flowchart2}, is introduced in NSA.
Therefore, a direct comparison based solely on the number of MCSs
with the results shown in Fig.~\ref{fig:G34_SQA} is not appropriate.

\subsection{\label{sec:level2}Results of NSA with Binary Spins}

\begin{figure}[t!]
\includegraphics[keepaspectratio,scale=0.64, bb=-20 0 500 627]{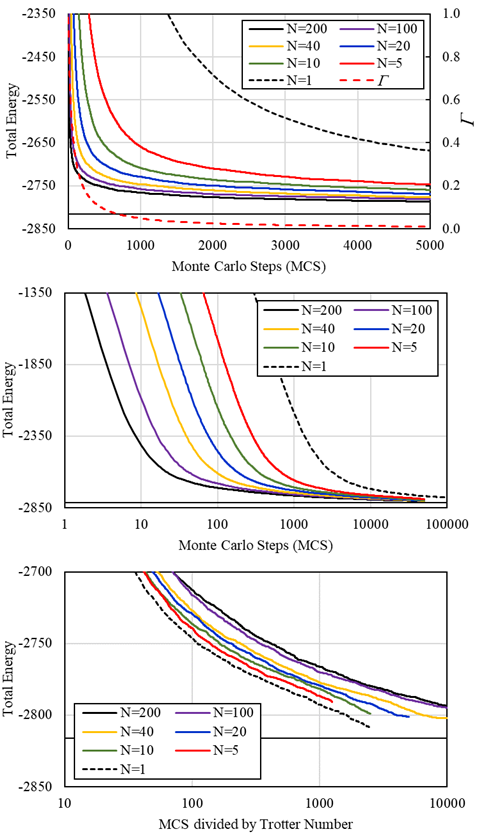}
\caption{
The results demonstrate that the partial optimization strategy inspired by causality can accelerate SQA
without requiring any additional hardware.
Since the product $N \hspace{-1.2pt} \times \hspace{-1.2pt} P$ is kept constant,
a smaller number of MCS required for convergence corresponds to a greater computational speedup.
As G34 is a sparse graph, the dependence on the number of trotter layers, $P$, is monotonic.
}
\label{fig:G34_3}
\end{figure}

\begin{figure}[t!]
\includegraphics[keepaspectratio,scale=0.64, bb=50 0 300 627]{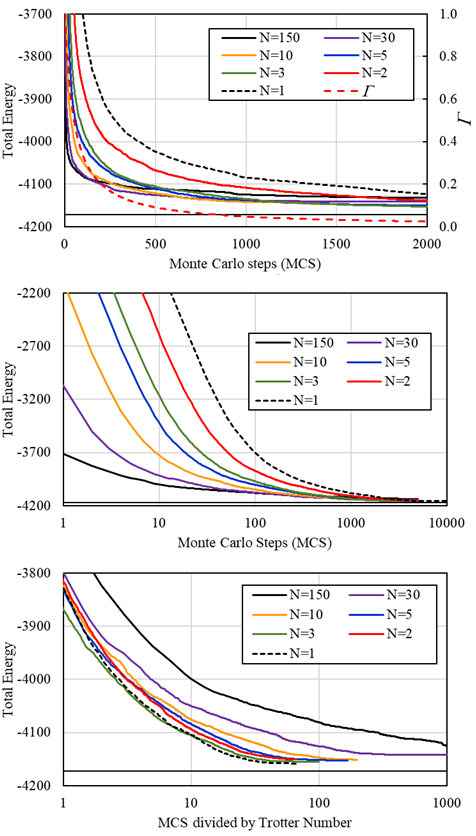}
\caption{
In comparison with G34, the effect of NSA is evident even for smaller values of $N$,
because G9 is a complex graph and the effect of a spin flip would propagates rapidly throughout the system.
However, the energy appears to saturate before reaching the global minimum energy.
This indicates that the search process is likely to be trapped in local minimum.
}
\label{fig:G9_3}
\end{figure}

\begin{figure}[t!]
\includegraphics[keepaspectratio,scale=0.62, bb=-20 0 500 575]{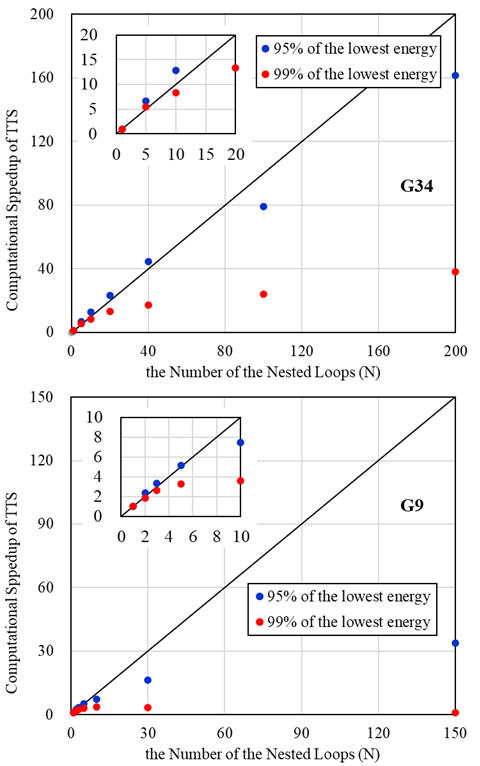}
\caption{
The TTS values required to reach 95\% and 99\% of the lowest energy are shown for the case of binary spins.
The speedup achieved with various numbers of nested loops is compared with that of NSA with $N=1$.
For the sparse graph G34, the acceleration factor is approximately proportional to 
$N$ for $N\hspace{-1pt}\leqq\hspace{-1pt}20$.
In contrast, for G9, the acceleration factor remains approximately proportional to 
$N$ only up to $N\hspace{-1pt}\leqq\hspace{-1pt}3$.
This difference is likely attributable to the graph characteristics.
}
\label{fig:tts}
\end{figure}

The dependence on the number of nested loops, $N$, is shown
in Fig.~\ref{fig:G34_3} for G34 and in Fig.~\ref{fig:G9_3} for G9 in the case of binary spins. 
The convergence behavior is presented in three different ways.
The upper graph shows the spin energy as a function of MCS on a linear scale,
whereas the middle graph shows the same data on a logarithmic scale.
It can be seen that increasing the number of nested loops reduces the number of MCSs required for convergence.
Fig.~\ref{fig:tts} clearly demonstrates that the Time to Solution (TTS) required to find the energy-minimum state
decreases as $N$. 
These results indicate that NSA can achieve computational speedup while requiring less memory.
Furthermore, when the trotter layers are processed in parallel,
an additional speedup could be achieved that is approximately the same TTS as the conventional SA.

In the bottom figure, the horizontal axis is rescaled by dividing the MCS by the trotter number, $P$.
This representation helps identify the most effective parameter settings.
For both G34 and G9, computational speedup is achieved as $N$ decreases,
corresponding to an increase in $P$.
However, in the case of G9, NSA achieves rapid convergence even with a small value of $P$.
As G9 is a complex graph, the effect of spin-flip propagates quickly to the whole system.
Therefore, NSA is thought to be more effective for complex optimization problems.

\begin{figure}[t!]
\includegraphics[keepaspectratio,scale=0.48, bb=-50 0 490 340]{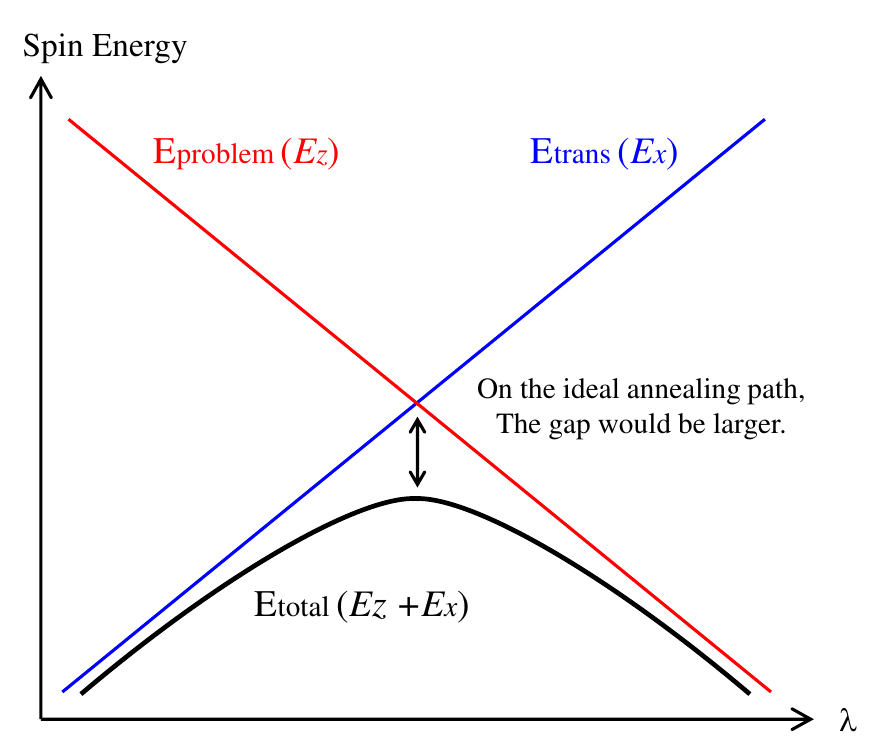}
\caption{
A conceptual illustration of the adiabatic phase transition is shown.
If QA is performed adiabatically, the system keep following the energy-minimum state.
Therefore, QA traces the adiabatic path with smaller energy in an ideal simulation.
So, the optimal annealing path may be the one in which the local maximum is minimized.
And the simulation needs fewer computational resources than the conventional SQA.
}
\label{fig:energy-crossing}
\end{figure}

\section{\label{sec:level1}Results of Continuous Variables}
\subsection{\label{sec:level2}Adiabatic Phase Transition}

In this section, spins are treated as continuous variables,
and it is shown that NSA effectively exploits local-minimum to find the global energy-minimum state.
Before presenting the results, I explain the conceptual diagram of the adiabatic phase transition
shown in Fig.~\ref{fig:energy-crossing}.
When $\lambda$ in Eq.~(7) is zero, all spins are aligned with the transverse electromagnetic field 
and remain in a trivial state.
As $\lambda$ gradually increases, each spin begins to direct in a more favorable direction
under the influence of the problem Hamiltonian.
If the change is sufficiently gradual, the system follows temporal energy-minimum state
and is expected to reach an eigenstate of the problem Hamiltonian eventually.
In addition, the local-maximum of $E_{z}$\hspace{0.03cm}+\hspace{0.03cm}$E_{x}$ represent the path
through which the system passes.
Therefore, NSAs with different parameters may have different local-maximum.
To test this hypothesis, $\lambda$ is controlled according to
$\lambda\hspace{0.03cm}$=$\hspace{0.03cm}1\hspace{-0.08cm}$ - $\hspace{-0.08cm}\mathit{\Gamma}$.

As a related research,
perturbative terms such as $X\hspace{-1pt}X$-catalysts have been actively studied~\cite{Wurburton_1,Wurburton_2,Matthias},
because positive off-diagonal matrix elements are associated with the ``sign problem''~\cite{Henelius_2000}.
Therefore, $X\hspace{-1pt}X$ interaction could accelerate SQA.
In those studies, the energy-gap change between the ground state and the first excited state was investigated
from fully quantum manners.
In my paper, I present the results the $X\hspace{-1pt}X$ interaction accelerates SQA by Monte Calro Simulation,
and discuss its mechanism in the next section.

\subsection{\label{sec:level2}Computational Results}

\begin{figure*}[t!]
\includegraphics[keepaspectratio,scale=0.98, bb=40 0 500 635]{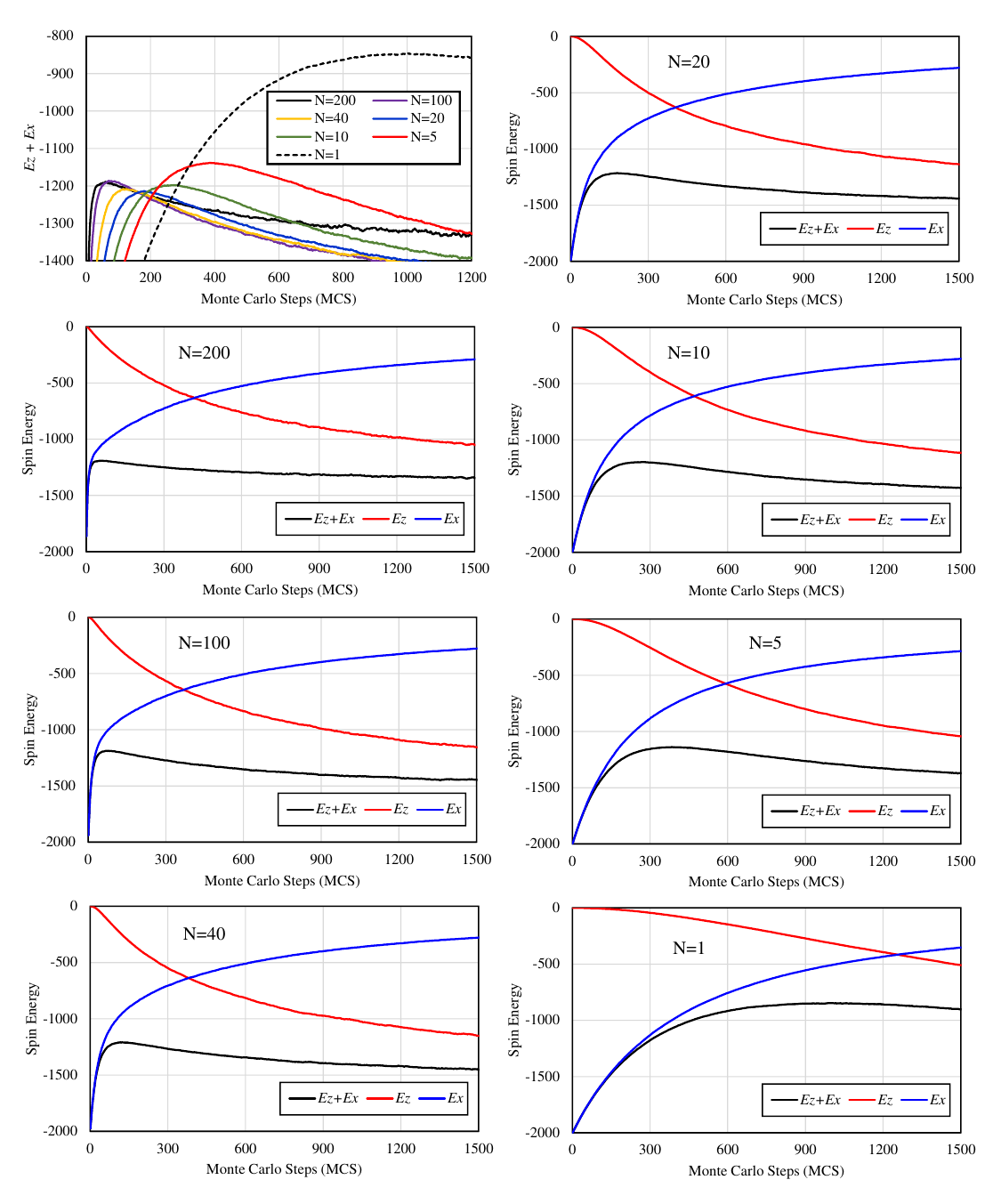}
\caption{
The upper-left graph shows the dependence of the total spin energy, $E_z$+$E_x$, on $N$.
The remaining seven graphs show $E_z$+$E_x$, $E_z$, and $E_x$ for various parameter near the local-maximum.
The quantity $E_z$+$E_x$ exhibits its smallest local-maximum near $N\hspace{-1pt}=\hspace{-1pt}20$,
which corresponds to using 10 trotter layers in the simulation.
In the case of NSA with $N\hspace{-1pt}\geq\hspace{-1pt}20$, 
the transverse electromagnetic field decreases so rapidly that the spins cannot adequately search for 
the energy-minimum state. 
Conversely, for NSA with $N\hspace{-1pt}\leq\hspace{-1pt}20$, the adiabatic phase transition can be realized.
Consistent with this interpretation, the acceleration ratio of NSA with $N\hspace{-1pt}\simeq\hspace{-1pt}10$ 
is the closest to linear scaling in Fig.~\ref{fig:tts}.
}            
\label{fig:G34_8}
\end{figure*}

As expected, both G34 and G9 have local-maximum for a certain parameter.
In the case of G34, when $N\hspace{0.03cm}$=$\hspace{0.03cm}1$, which corresponds to the conventional SQA
with the trotter number, $P\hspace{0.03cm}$=$\hspace{0.03cm}200$, 
the largest local-maximum is obtained, and the phase transition is considered to be the most adiabatic.
This behavior is shown in the upper-left graph of Fig.~\ref{fig:G34_8}.
On the other hand, when $N\hspace{0.03cm}$=$\hspace{0.03cm}200$, 
the spins tend to remain aligned with the transverse electromagnetic field 
before orienting themselves toward their favorable directions.
This behavior is clearly indicated by a rapid increase in $E_{z}$\hspace{0.03cm}+\hspace{0.03cm}$E_{x}$
at small MCS values. 
The smallest local-maximum appears around $N\hspace{0.03cm}$=$\hspace{0.03cm}20$.
Such annealing path is considered to be the most efficient 
because the energy increase caused by the transverse electromagnetic field is suppressed 
and the required computational resources are reduced to one-tenth.
The result shown in Fig.~\ref{fig:tts}, where the TTS required to reach 95\% of the global minimum energy is accelerated 
almost proportionally up to $N\hspace{0.03cm}$=$\hspace{0.03cm}20$, supports this interpretation.

\begin{figure*}[t!]
\includegraphics[keepaspectratio,scale=0.98, bb=40 0 500 635]{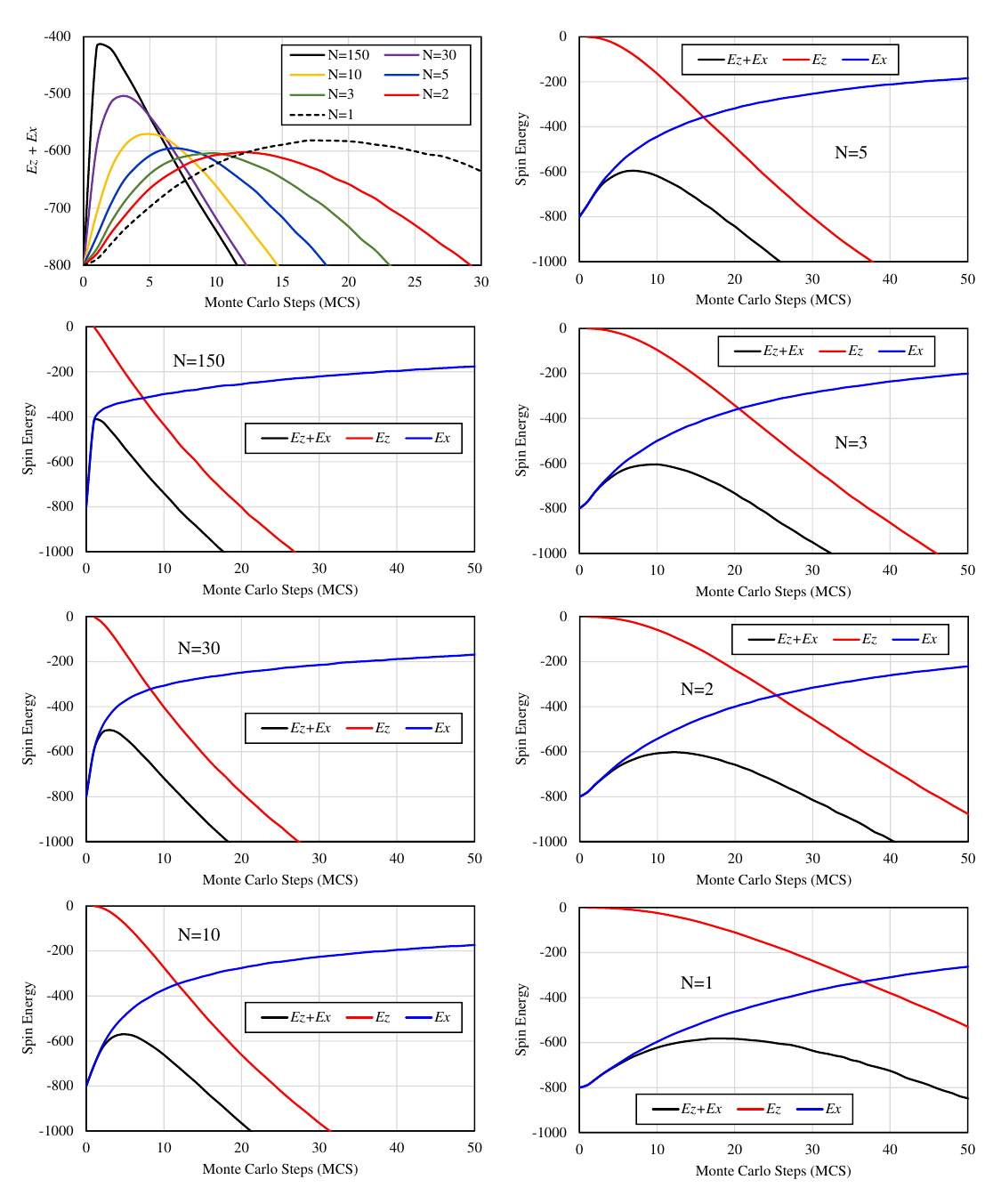}
\caption{
As in Fig.~\ref{fig:G34_8}, the upper-left graph shows the dependence of the total spin energy, $E_z$+$E_x$, on $N$. 
The remaining graphs show $E_z$+$E_x$, $E_z$, and $E_x$ for various parameter. 
The upper-left graph indicates that $E_z$+$E_x$ exhibits its smallest local-maximum near $N\hspace{-1pt}=\hspace{-1pt}3$,
which corresponds to using 50 trotter layers in the simulation. 
In the case of NSA with $N\hspace{-1pt}\leq\hspace{-1pt}3$, the adiabatic phase transition can be realized.
The convergence for $N\hspace{-1pt}=\hspace{-1pt}3$ is faster than that for any other value, 
as shown in the bottom graph of Fig.~\ref{fig:G9_3}.
}
\label{fig:G9_8}

\end{figure*}

Similarly, Fig.~\ref{fig:G9_8} shows that a local-maximum appears near $N\hspace{0.03cm}$=$\hspace{0.03cm}3$ for G9.
The bottom graph in Fig.~\ref{fig:G9_3} also indicates that the NSA with $N\hspace{0.03cm}$=$\hspace{0.03cm}3$ 
converges faster than any other NSA until $E_{z}$\hspace{0.03cm}=\hspace{-0.1cm} -- \hspace{-0.1cm}$4,000$.
Furthermore, Fig.~\ref{fig:tts} shows that the TTS is accelerated approximately in proportion to $N$ 
in the range $N\hspace{-1pt}\leqq\hspace{-1pt}3$.
These results support the hypothesis that the NSA with smaller local-maximum follows
both an adiabatic and an hardware-efficient annealing path.

\subsection{\label{sec:level2}Merit of NSA with Continuous Variables}

The results presented above clearly demonstrate that the proposed method is useful for selecting annealing parameters, 
because it only requires identifying the smallest local-maximum.
When solving a QUBO problem, choosing appropriate annealing parameters is often difficult, 
since the optimization process must be executed repeatedly until completion.
However, the proposed method can assist in parameter selection with far fewer MCSs, 
because the local-maximum can be easily found during the annealing process.
Although this paper presents the results for the number of nested loops, 
the same approach can be applied to determine other annealing parameters, such as $T$ and $\mathit{\Gamma}$.

\section{\label{sec:level1}Reasons for Acceleration of NSA}
\subsection{\label{sec:level2}The Origin of $X\hspace{-1pt}X$ interaction}

In this section, the role of the $X\hspace{-1pt}X$ interaction is reconsidered, 
and its connection to the UCC method~\cite{Troyer} is clarified.
In quantum chemistry, the accuracy of ground-state energy calculations can be improved 
using the UCC method.
In this approach, possible single- and double-electron excitations are taken into account 
to incorporate the correlation energy neglected in the 
Hartree\hspace{-0.1cm}-- \hspace{-0.1cm}Fock approximation. 
Furthermore, higher-energy states decay exponentially faster 
under Quantum Imaginary Time Evolution (QITE)~\cite{Kubra,Fan}. 

Applying this idea to optimization problems, we note that, 
in the Ising model, the state obtained by flipping all spins is degenerate with the original state
and therefore has the same energy.
In fact, when the state is expressed as $|\psi_0\rangle = \bigl ( \cos(\frac{\theta^{i'}}{2}) , \sin(\frac{\theta^{i'}}{2}) \bigr ) 
\otimes \bigl ( \cos(\frac{\theta^{j'}}{2}) , \sin(\frac{\theta^{j'}}{2}) \bigr ) $,
the expectation value $\langle \psi_0 |  \hat{\sigma}_{z}^{i'} \hat{\sigma}_{x}^{i'} \hat{\sigma}_{z}^{j'} \hat{\sigma}_{x}^{j'} | \psi_0 \rangle$ is vanishes.
Therefore, the spin energy can be calculated as below:

\small
\begin{align}
    E             &= \langle \psi_0 | e^{-H \tau}  H e^{-H \tau} 
                                     | \psi_0 \rangle \nonumber \\
                    &= \Bigl \langle \psi_0 \Bigl | e^{-H \tau}  H 
                            \exp \Bigl(-\bigl(H - \sum_{i'<j'} J_{i'j'} \hat{\sigma}_{z}^{i'} \hat{\sigma}_{x}^{i'} \hat{\sigma}_{z}^{j'} \hat{\sigma}_{x}^{j'} \bigr) \tau \Bigr)
                                    \Bigr | \psi_0 \Bigr \rangle. \nonumber
\end{align}
\normalsize

Here, $H$ denotes the first term of Eq.~(1), 
and the fact that the two terms commute is used,
Therefore, the energy of the system can be calculated as follows:

\small
\begin{align}
    E             
                   &= \Bigl \langle \psi_0 \Bigl | e^{-H \tau}  H e^{- H \tau} 
                            \exp \Bigl(+\sum_{i'<j'} J_{i'j'} \hat{\sigma}_{z}^{i'} \hat{\sigma}_{x}^{i'} \hat{\sigma}_{z}^{j'} \hat{\sigma}_{x}^{j'} \tau \Bigr) \Bigr | \psi_0 \Bigr \rangle \nonumber \\
                   &= \Bigl \langle \psi_0 \Bigl | e^{-2H \tau} H  
                            \exp \Bigl(+\sum_{i'<j'} J_{i'j'} \hat{\sigma}_{z}^{i'} \hat{\sigma}_{x}^{i'} \hat{\sigma}_{z}^{j'} \hat{\sigma}_{x}^{j'} \tau \Bigr) \Bigr | \psi_0 \Bigr \rangle \nonumber
\end{align}
\normalsize

By approximating the exponential term to first order and 
keeping the terms up to the second-order Pauli product,
the energy expression can be simplified to

\small
\begin{align}
    E             &= e^{-2E \tau} \Bigl \langle \psi_0 \Bigl |   
                            - \sum_{i<j} J_{ij} \hat{\sigma}_{z}^{i} \hat{\sigma}_{z}^{j}
                            \Bigl(I + \hspace{-0.1cm}\sum_{i'<j'} J_{i'j'} \hat{\sigma}_{z}^{i'} \hat{\sigma}_{x}^{i'} \hat{\sigma}_{z}^{j'} \hat{\sigma}_{x}^{j'} \tau \Bigr) \Bigr| \psi_0 \Bigr \rangle \nonumber \\
                   &= e^{-2E \tau} \Bigl \langle \psi_0 \Bigl |   
                            - \sum_{i<j} J_{ij} \hat{\sigma}_{z}^{i} \hat{\sigma}_{z}^{j} \nonumber \\
                   & \hspace{2.5cm} - \sum_{i<j} \sum_{i'<j'} J_{ij} J_{i'j'} \hat{\sigma}_{z}^{i} \hat{\sigma}_{z}^{j} \hat{\sigma}_{z}^{i'} \hat{\sigma}_{x}^{i'} \hat{\sigma}_{z}^{j'} \hat{\sigma}_{x}^{j'} \tau \Bigr|
\psi_0 \Bigr \rangle \nonumber \\
                  &= e^{-2E \tau} \Bigl \langle \psi_0 \Bigl |   
                            - \sum_{i<j} J_{ij}    \hat{\sigma}_{z}^{i} \hat{\sigma}_{z}^{j}
                            - \sum_{i<j} J_{ij}^{\hspace{0.05cm}2} \hat{\sigma}_{x}^{i} \hat{\sigma}_{x}^{j} \tau \Bigr| \psi_0 \Bigr \rangle \nonumber \\
                   &= e^{-2E \tau} \Bigl \langle \psi_0 \Bigl |   
                            - \sum_{i<j} J_{ij} \hat{\sigma}_{z}^{i} \hat{\sigma}_{z}^{j}
                            - \sum_{{\mathcal{E}_\mathit{ij}}} \hat{\sigma}_{x}^{i} \hat{\sigma}_{x}^{j} \tau \Bigr| \psi_0 \Bigr \rangle. 
\end{align}
\normalsize

Based on the resulting expression for the energy in Eq.~(14), 
the corresponding Hamiltonian can be derived as

\small
\begin{align}
    H  = - \sum_{i<j} J_{ij} \hat{\sigma}_{z}^{i} \hat{\sigma}_{z}^{j}
                  - \sum_{{\mathcal{E}_\mathit{ij}}} \hat{\sigma}_{x}^{i} \hat{\sigma}_{x}^{j} \tau.
\end{align}
\normalsize

When deriving Eq.~(15), we use the fact that $J_{ij}^{\hspace{0.05cm}2}=1$, 
since $J_{ij}=\pm1$ for both G9 and G34, 
and that the $X\hspace{-1pt}X$ interaction is nonzero only when the edge connecting the $i$-th and $j$-th nodes, ${\mathcal{E}}_{ij}$, exists. 
Furthermore, since the Hamiltonian of the system should remain unchanged 
after the perturbation, the last term must vanish. 
Therefore, $\tau$ in Eq.~(15) should be chosen to be $s(1\hspace{0.1cm}$--$\hspace{0.1cm}s)$~\cite{Wurburton_2}.

\subsection{\label{sec:level2}Applicability to Other Non-Stoquastic Term}

In my simulation presented in this paper, the Python code shown below is used.

When the angle of the $i$-th spin on the $k$-th trotter layer is updated ($\theta_b^i \rightarrow \theta_a^i$), 
the spins that directly interact with it (the $j$-th spins) are considered, 
and all corresponding energy changes are summed. 
And, the next spin is preferentially selected from among the spins adjacent to the $i$-th spin, 
as shown in Fig.~\ref{fig:flowchart2}. 
Since the choice of the spin involved in the $X\hspace{-1pt}X$ interaction is arbitrary,
the spin can be selected at random.
Therefore, this selection process should be repeated among the spin groups.
This procedure results in ``partial optimization'' and it is reasonable because NSA is based on the causality.

\begin{figure}[h!]
\includegraphics[keepaspectratio,scale=0.55, bb=-10 0 490 55]{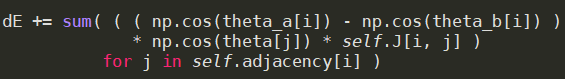}
\label{fig:dE}
\end{figure}

As explained above, the $X\hspace{-1pt}X$ interaction is originated from quantum mechanical approach of UCC
and effectively utilizes the degeneracy of the system. 
Therefore, although NSA is performed on a classical computer by updating two spins, 
the algorithm can be thought to exploit the quantum effect.
If the idea making use of UCC can be extended to the other non-stoquastic Hamiltonian
and it can be reduced to the Pauli-product that is classically computable,
additional computational speedup could be achieved on Monte Carlo Simulation.
Although this is a highly challenging task, it is worth investigating.

\section{\label{sec:level1}Conclusions}

A new SQA method, in which the trotter layers are handled in parallel, is proposed. 
In this algorithm, the quantum effect is retained as a penalty term to encourage all spins to align with 
those in the temporally energy-minimum state.
Although the method is not mathematically rigorous,
it tends to converge faster than the conventional SA and SQA.

Next, I propose a new heuristic approach, NSA. 
In this algorithm, a nested loop that functions as a form of ``partial optimization'' 
is able to accelerate convergence without requiring any additional hardware resources. 
Furthermore, the results show that the convergence speed could be improved further 
when multiple trotter layers are handled in parallel. 
This suggests that both quantum and classical effects contribute effectively to the optimization process, 
since the trotter layers originate from quantum mechanics. 

Finally, I formulate the total spin energy for the case in which spins are treated as continuous variables. 
Through simulations, it is shown that the convergence speed has a clear relationship 
with the local-maximum of the spin energy. 
I also demonstrate that suitable annealing parameters could be selected efficiently 
by determining whether the system passes through the adiabatic phase transition. 
These results suggest that the proposed method is effective for solving QUBO problems on classical computers.
Furthermore, the origin of the $X\hspace{-1pt}X$ interaction is investigated using the UCC method
which has been successfully applied in quantum chemistry, 
and discuss the possibility of Monte Carlo Simulation containing non-stoquastic Hamiltonians. 
The approach that I propose may allow us to treat the problems 
that are generally considered difficult for classical computers
and could help solve practical optimization problems more easily.

\end{document}